\documentclass[shortnote,twocolumn]{jpsj3}
\usepackage{txfonts}
\usepackage{color}
\usepackage{ulem}

\title{Site-Selective Antimony Doping in Arsenic Zigzag Chains of 112-type Ca$_{1-x}$La$_x$FeAs$_2$}

\author{
Hiromi Ota$^1$, 
Kazutaka Kudo$^2$\thanks{kudo@science.okayama-u.ac.jp}, 
Takumi Kimura$^2$, Yutaka Kitahama$^2$, 
Tasuku Mizukami$^2$, Satoshi Ioka$^2$, 
and Minoru Nohara$^2$
}

\inst{
$^1$Advanced Science Research Center, Okayama University, Okayama 700-8530, Japan \\
$^2$Research Institute for Interdisciplinary Science, Okayama University, Okayama 700-8530, Japan \\
} 

\abst{
Single crystal X-ray diffraction studies were performed for the Sb-doped 112-type iron-based superconductor Ca$_{1-x}$La$_x$FeAs$_2$ with the superconducting transition temperature $T_{\rm c}$ of 47 K. 
Doped Sb preferably substituted not for As(1) in the FeAs layers but for As(2) in the layers of As zigzag chains. 
Structural reasons for $T_{\rm c}$ enhancement by Sb doping were discussed. 
}

\begin{document}
\maketitle

Recently, novel 112-type iron-based superconductors have received considerable attention\cite{Katayama_2013,Yakita_2014,Sala_2014} because the superconductivity induced by the simultaneous doping\cite{Kudo_SciRep} of rare earth ($RE$) elements and antimony results in superconducting transition temperatures up to 47 K\cite{Kudo_2014_1,Kudo_2014_2}.

The chemical formula of these 112-type iron-based materials is written as Ca$_{1-x}$$RE_{x}$FeAs$_2$ ($RE$ = La, Ce, Pr, Nd, Sm, Eu, Gd)\cite{Katayama_2013,Yakita_2014,Sala_2014} because the partial substitution of $RE$ for Ca is necessary to stabilize the 112 phase\cite{Katayama_2013,Yakita_2014,Sala_2014}.
These materials crystalize in a monoclinic structure with space group $P2_1$ (No. 4, $C_2^2$)\cite{Katayama_2013,Harter_2016}. 
The structure consists of alternately stacked layers of FeAs and As zigzag chains, as shown in Fig. 1. 
In other words, there are two As sites: As(1) in the FeAs layers, and As(2) in the layers of zigzag chains. 
Hence, the chemical formula can also be written as [Ca$_{1-x}$$RE_x$][FeAs(1)][As(2)].

Most 112-type iron-based materials exhibit superconductivity at 10--35 K\cite{Katayama_2013,Yakita_2014,Sala_2014,Kudo_2014_1,Kudo_2014_2}. 
Among these materials, the sample with $RE$ = La exhibited superconductivity at 35 K\cite{Katayama_2013,Kudo_2014_1,Kudo_2014_2}.  
With Sb doping, $T_{\rm c}$ is drastically increased to 47 K\cite{Kudo_2014_2}, which is the highest $T_{\rm c}$ among 112-type iron-based materials. 
This result markedly contrasted the less-pronounced Sb-doping effect in 1111-type LaFeAsO$_{1-x}$F$_x$\cite{Carlsson_2011,Cao_2010,Singh_2009}. 
In 112-type iron-based materials, one remaining issue is understanding the preferential As site for Sb dopants. 
First-principles calculations have predicted that the substitution of Sb for As(2) in the zigzag chains is more stable than that for As(1) in the FeAs layers\cite{Nagai_2015}. 
However, this theoretical prediction has not been experimentally verified.

In this paper, we report the results of single-crystal X-ray structure analysis for 112-type Ca$_{1-x}$La$_{x}$Fe(As$_{1-y}$Sb$_y$)$_2$. 
The results suggested that Sb is selectively doped into the As(2) site in the zigzag chains.

Single crystals of Ca$_{1-x}$La$_{x}$Fe(As$_{1-y}$Sb$_y$)$_2$ were grown by heating a mixture of Ca, La, FeAs, As, and Sb powders with nominal compositions of $x$ = 0.10 and $y =$ 0.10. 
The details of the crystal-growth process have been previously reported\cite{Kudo_2014_1,Kudo_2014_2}. 
Following the crystal-growth process, plate-like single-crystalline samples of Ca$_{1-x}$La$_{x}$Fe(As$_{1-y}$Sb$_y$)$_2$ were obtained together with a powder mixture of LaAs, FeAs, FeAs$_2$, and CaFe$_2$As$_2$. 
Single crystals with typical dimensions of 0.03 $\times$ 0.03 $\times$ 0.01 mm$^3$ were used for single-crystal X-ray diffraction (XRD) experiments which were performed using Rigaku Single Crystal X-ray Structural Analyzer (Varimax with Saturn). 
For the same batch of samples, we confirmed the emergence of bulk superconductivity at 47 K using several single crystals with typical dimensions of 0.4 $\times$ 0.4 $\times$ 0.02 mm$^3$ by measuring the temperature dependence of magnetization with a magnetic property measurement system (MPMS; Quantum Design). 
\begin{figure}[t]
\begin{center}
\includegraphics[width=5cm]{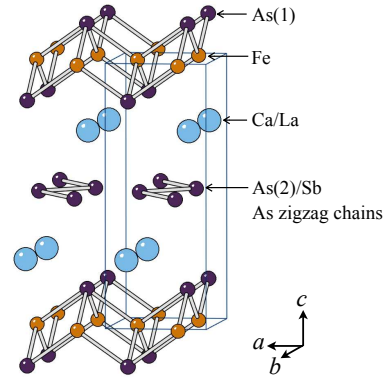}
\caption{
(Color online) Crystal structure of Ca$_{1-x}$La$_{x}$Fe(As$_{1-y}$Sb$_y$)$_2$ that crystallizes in a monoclinic structure with space group $P2_1$ (No. 4, $C_2^2$). 
Here, As(1) and As(2) correspond to the arsenic sites in the FeAs layers and the layers of As zigzag chains, respectively. 
Solid lines indicate the unit cell. 
} 
\end{center}
\end{figure}
\begin{table}[t]
\caption{Data collection and refinement statistics for the single-crystal X-ray structure analysis of Ca$_{1-x}$La$_{x}$Fe(As$_{1-y}$Sb$_{y}$)$_2$ at 110 K.}
\begin{center}
\begin{tabular}{ll}
\hline
Crystal System 				& monoclinic \\
Space Group 				& $P2_1$ (No. 4, $C_2^2$)\\
$a$ (\AA) 					& 3.983(17) \\
$b$ (\AA) 					& 3.939(16) \\
$c$ (\AA) 					& 10.35(4) \\
$\beta$ ($^{\circ}$) 			& 90.74(5) \\
Volume (\AA$^3$) 			& 162.4(11) \\
$R_{\rm merge}$ (\%) 		& 8.93 \\
Completeness (\%) 			& 99.7 \\
Z value 					& 2\\
&\\
Radiation type 				& Mo $K\alpha$ \\
Radiation wave length (\AA) 	& 0.71075 \\
No. of Reflections 			& 2318 \\
$R$1 (\%) 					& 9.14\\
\hline
\end{tabular}
\end{center}
\end{table}
\begin{table}[t]
\caption{
The occupancies, atomic coordinates, and equivalent isotropic atomic displacement parameters are listed for Ca$_{1-x}$La$_{x}$Fe(As$_{1-y}$Sb$_{y}$)$_2$ [$x =$ 0.100(18), $y =$ 0.06(4)] that crystallizes in a monoclinic structure with space group $P2_1$ (No. 4, $C_2^2$) at 110 K. 
}
\begin{center}
\begin{tabular}{p{2.9em}p{3.7em}p{3.7em}p{3.7em}p{4.3em}p{3em}}
\hline
       &                    &           &           &           & 100$U_{\rm eq}$\\
\hfil Site \hfil & \hfil Occ. \hfil & \hfil $x/a$ \hfil & \hfil $y/b$ \hfil & \hfil $z/c$ \hfil & \hfil (\AA$^2$) \hfil\\
\hline
As(1)/Sb & 0.99(5) & 1.2541(8) & 0.6538(15) & 0.1360(3) & 1.41(11)\\
               & /0.01(5) & & & &\\
As(2)/Sb & 0.88(5)  & 0.7267(10) & 0.667(2) & 0.5032(3) &2.29(14)\\
               & /0.12(5) & & & &\\
Ca/La      & 0.900(18) & 0.7583(13) & 1.162(3) & 0.2665(5) & 1.5(2)\\ 
                & /0.100(18) & & & &\\
\hfil Fe \hfil           & \hfil 1 \hfil           & 1.2484(12) & 0.159(5) & -0.0006(5) & 1.55(16)\\
\hline
\end{tabular}
\end{center}
\end{table}

The single-crystal XRD data of the Ca$_{1-x}$La$_{x}$Fe(As$_{1-y}$Sb$_{y}$)$_2$ samples were refined based on its monoclinic structure with space group $P2_1$ (No. 4, $C_2^2$), as shown in Tables I and II. 
The unweighted agreement factor $R1$ is sufficiently small to conclude that the analysis was well converged. 
As shown in Table II, the occupancies, atomic coordinates, and equivalent isotropic atomic displacement parameters for all constituent elements were successfully refined.

The refined structural parameters suggest that Sb atoms selectively substitute for As atoms in the zigzag chains. 
The occupancies of Sb at the As(1) and As(2) sites were estimated to be 0.01(5) and 0.12(5), respectively. 
The occupancy at As(2) is larger than that at As(1) in the entire range of the error, suggesting selective doping of Sb. 
At this stage, it is not clear whether all Sb atoms are doped into As(2) site; in other words, the results do not show whether Sb atoms substitute for As(1). 
These results are consistent with the prediction from the first principles calculations\cite{Nagai_2015}.

The chemical composition suggested from the refined parameters is Ca$_{1-x}$La$_{x}$Fe(As$_{1-y}$Sb$_{y}$)$_2$ [$x =$ 0.100(18), $y =$ 0.06(4)].  
The Sb content is found to be slightly lower than the nominal one. 
In the present system, the Sb content could not be determined by energy-dispersive X-ray spectrometry (EDS)\cite{Kudo_2014_2} because the Sb peak positions in the EDS spectra overlapped those of Ca.

In Ref. \citen{Kudo_2014_2}, we reported the preliminary results of the structure analysis in the present system. 
The analysis could not determine the structural parameters for doped Sb, but it provided a preliminary suggestion that the increase in the $b$ parameter due to Sb doping makes the As-Fe-As angle $\alpha$ along the $b$-axis, which is distorted in the monoclinic structure, closer to the ideal value ($\alpha =$ 109.47$^\circ$) at which iron-based superconductors generally tend to exhibit high $T_{\rm c}$\cite{Lee_2008}. 
The present study confirms the suggestion. 
In the present sample, the $a$ and $b$ parameters are increased compared with the lattice parameters of the Sb-free samples \cite{Katayama_2013,Kudo_2014_1,Kudo_2014_2}, while the $c$ parameter is hardly changed. 
The angles $\alpha_a$ and $\alpha_b$, which correspond to $\alpha$ along the $a$- and $b$-axes, respectively, were estimated to be 109.7(4)$^\circ$ and 108.7(4)$^\circ$ in the present sample, while $\alpha_a =$ 109.099(9)$^\circ$ and $\alpha_b =$ 107.061(9)$^\circ$ in the Sb-free sample\cite{Katayama_2013}. 
The improvement in $\alpha_b$ due to the increase in the $b$ parameter is also observed in the present analysis.

Finally, we discuss the importance of selective Sb doping into the As zigzag chains in the present material. 
In general, chemical doping in iron-based superconductors are categorized into ``direct doping" and ``indirect doping", where direct doping is doping into the Fe site, while indirect doping is the doping into sites other than the Fe site\cite{Katase_2013}. 
Although superconductivity in iron-based materials is basically robust against doping-induced disorder\cite{Hosono_2015}, indirect doping\cite{Ren_2008,Hanna_2011,Matsuishi_2014} appears to be more effective than direct doping\cite{Qi_2008} in achieving high $T_{\rm c}$, which might be due to reduced structural perturbation in the superconducting layers. 
For example, $T_{\rm c}$ = 55 K in SmFeAsO$_{1-x}$F$_x$\cite{Ren_2008}, 55 K in SmFeAsO$_{1-x}$H$_x$\cite{Hanna_2011}, and 36 K in SmFeAs$_{1-y}$P$_y$O$_{1-x}$H$_x$\cite{Matsuishi_2014}, while $T_{\rm c}$ = 15.2 K in SmFe$_{1-x}$Co$_x$AsO\cite{Qi_2008}. 
In the present material, the Sb doping is considered indirect doping\cite{RE_Ca}. 
Moreover, the doped Sb preferably substitutes not for As(1) in the FeAs layers but for As(2) in the layers of As zigzag chains; hence, the Fe site is largely uninfluenced by doping-induced disorder. 
The situation is compatible with the emergence of high $T_{\rm c}$. 
On the other hand, the Sb doping must induce substantial disorder into the As zigzag chains. 
First-principles calculations have proposed that the ${p}_{x}$ and ${p}_{y} $ orbitals of As(2) in the unique zigzag structure would result in an anisotropic Dirac cone near the Fermi level, which is gapped by the spin-orbit coupling and is topologically nontrivial\cite{Wu_2014,Wu_2015}. 
The Dirac-cone-like band structure was confirmed by recent angle-resolved photoemission spectroscopy experiments\cite{Liu_2016}.
Thus, Sb-doped 112-type materials may provide a unique opportunity for studying disorder effects on the Dirac cone.

In summary, we investigated the Sb site in 112-type Ca$_{1-x}$La$_{x}$Fe(As$_{1-y}$Sb$_y$)$_2$ with $T_{\rm c}$ of 47 K. 
The single-crystal X-ray structure analysis suggested that the doped Sb atoms preferably substitute for As atoms in the layers of zigzag chains. 
The Sb-doped As zigzag chains may become a useful platform for investigating the effects of disorder on topologically nontrivial band structures.

\acknowledgment
This work was partially supported by Grants-in-Aid for Scientific Research (No. 26287082, 15H01047, 15H05886, and 16K05451) provided by the Japan Society for the Promotion of Science (JSPS) and the Program for Advancing Strategic International Networks to Accelerate the Circulation of Talented Researchers from JSPS. 


\begin{thebibliography}{99}
\bibitem{RE_Ca} In the present system, the doping of rare earth elements into the Ca site is also considered indirect doping.
\end{thebibliography}
\end{document}